# Foundational Issues in Group Field Theory

Álvaro Mozota Frauca[1]

**Abstract**
In this paper I offer an introduction to group field theory (GFT) and to some of the issues affecting the foundations of this approach to quantum gravity. I first introduce covariant GFT as the theory that one obtains by interpreting the amplitudes of certain spin foam models as Feynman amplitudes in a perturbative expansion. However, I argue that it is unclear that this definition of GFTs amounts to something beyond a computational rule for finding these transition amplitudes and that GFT doesn't seem able to offer any new insight into the foundations of quantum gravity. Then, I move to another formulation of GFT which I call canonical GFT and which uses the standard structures of quantum mechanics. This formulation is of extended use in cosmological applications of GFT, but I argue that it is only heuristically connected with the covariant version and spin foam models. Moreover, I argue that this approach is affected by a version of the problem of time which raises worries about its viability. Therefore, I conclude that there are serious concerns about the justification and interpretation of GFT in either version of it.

Group field theory (GFT) is an approach to quantum gravity that has attracted some attention recently.[1] In this paper I offer an introduction to the approach and I point to some issues affecting its foundations that in my opinion jeopardize the viability of the approach. In particular, I will argue that even if it is spoken about group field theories as a single approach we can distinguish between two types of approaches, which I will refer to as covariant and canonical. These two approaches are assumed to be equivalent, but I will argue against this claim. Furthermore, I will argue that both perspectives on group field theory have major conceptual shortcomings.

---

[1] See for instance [1, 2].

✉ Álvaro Mozota Frauca
  alvaro.mozota@udg.edu

1  Department of Physics, University of Girona, Carrer de la Univeristat de Girona, 1, Girona 17003, Girona, Spain

   Springer

I will start in Sect. 1 by giving an introduction to GFT and by giving a motivation for adopting this formalism that comes from several approaches to quantum gravity. These approaches have some common structures, and GFT has been argued to capture these common features and to offer an improvement over them.[2] In particular, I will focus on spin foam models and I will show how they naturally arise in the perturbative expansions of GFTs as Feynman diagrams. In this sense, GFTs can be seen as a useful way of computing quantities defined in spin foam models and as an improvement over them, as they provide a guideline about how to sum over different spin foams. The relationships between spin foams and GFT, spin foams and other approaches to quantum gravity, and GFT itself with these approaches can be argued to support the adoption of GFT as a useful approach to quantum gravity.

In Sect. 2 I introduce covariant GFTs precisely as the view of GFTs which is more closely related to spin foam models and which is associated with covariant quantizations, i.e., with quantizations relying on a path integral structure. I will argue that when one studies the foundations of the approach, one reaches the conclusion that it is unclear what this approach amounts to and to what extent this constitutes a quantum theory.

Finally, in Sect. 3 I discuss an alternative approach to GFT which is based on a formalism similar to the canonical formalism of quantum mechanics, i.e., it is based on a Hilbert space structure. In particular, this approach mimics the canonical version of loop quantum gravity (LQG) and defines a kinematical Hilbert space which is a Fock space but which is very similar to the spin network Hilbert space of LQG. The dynamics of this version of GFT is also defined in a similar way to LQG, that is, physical states are the ones that are part of a physical Hilbert space which is defined by imposing some constraint on the kinematical Hilbert space. In the case of LQG this constraint corresponded to the Hamiltonian constraint of general relativity, while in the case of GFT this constraint is motivated by the GFT action. Despite this difference, I will argue that canonical GFT suffers from a problem of time as LQG and other canonical approaches to quantum gravity. In particular, in the GFT context a relational strategy for dealing with the problem is popular and I argue that it introduces time in an ad hoc and unsatisfactory way and that this raises worries about the feasibility of the approach.

## 1 Spin Foam Models as Hinting to GFT

In the context of quantum gravity several approaches have shown a certain degree of convergence in the last decades. Approaches like loop quantum gravity, causal dynamical triangulations, quantum Regge calculus, or matrix models have converged to similar discrete structures. In particular, spin foams are a structure that can be related to several of them, and group field theory can be motivated from the perspective of this family of models. In this section I will focus on the relationship between spin foam models and GFT and I will refer the reader to the literature[3] for more detailed accounts of how GFT and spin foam models relate to other approaches to quantum gravity.

---

[2] See for instance [3].

The most straightforward way of thinking about spin foams is as dual to a triangulation of spacetime.[4] In several approaches to quantum gravity it is postulated that spacetime is in some sense discrete or made of chunks. For our discussion I will take these discrete pieces to be flat and the simplest available: for 2-dimensional spaces this corresponds to triangles; in 3 dimensions, tetrahedra; and in 4-dimensions, 4-simplices, which are just the generalization of triangles to higher dimensions. Given any triangulation of spacetime, we can build a dual structure that encodes the relations between its different pieces. This dual structure is known as 2-complex, and it is made of 2-dimensional surfaces, 1-dimensional edges, and 0-dimensional vertices that represent the structure of the triangulation. For instance, in three dimensions every vertex represents a tetrahedron, an edge joining two vertices represents that the tetrahedra associated with them share a triangle and a surface of the 2-complex represents a segment of the triangulation. This is represented in Fig. 1 and contained in Table 1

A spin foam is obtained by 'coloring' the 2-complex, i.e., by assigning the appropriate quantities to its surfaces and edges. In particular, these quantities correspond to holonomies of the gravitational connection and fluxes of forms which roughly correspond to geometric properties such as areas of triangles for the 4-dimensional case. In this sense, the 2-complex gives the basic relations of the discrete geometry and by coloring it we get the full geometry.

A spin foam model is basically a path integral defined on a 2-complex. That is, in the same way that in a path integral one sums over all possible configurations of the field between an initial and a final moment of time, in a spin foam model one sums over all possible geometries that a 2-complex can have. In this case, the initial and final configurations correspond to discrete geometries of space, which are represented by spin networks and the transition amplitude $K$ represents something like the probability amplitude of observing a final discrete geometry given that an initial one has been observed. For our discussion here it will be enough to notice that these amplitudes usually take a form like the following[5]

$$K(h_l) = \mathcal{N} \int dh_{vf} \prod_f \delta(h_f) \prod_v A(h_{vf}). \qquad (1)$$

Here one is summing over all possible colorings of the 2-complex compatible with the boundary spin network, that is with all the assignments of the internal holonomies $h_{vf}$ compatible with the external ones $h_l$. $\mathcal{N}$ is some normalization factor, $h_f$ corresponds to a product of holonomies associated with each face of the 2-complex and *A* is a function associated with each vertex of the 2-complex which depends on the holonomies in such a vertex. I refer the reader to [4] for more details and a justification of this expression for the EPRL spin foam model, although this expression is applicable for other spin foam models, which would differ on some details or the exact form of the vertex amplitude *A*.

---

[3] For an introduction to spin foam models and their relation with other approaches see [4], and for an account of how GFT can be seen as the converging point of these approaches see [3].
[4] This is not the only way of thinking about them, and in certain approaches there are spin foams which cannot be understood as dual to triangulations. See [5] for a general definition of spin foams.
[5] This is, for instance, the case of the EPRL model as introduced in Ref. [4], both for the 3-dimensional case [4, p. 114] and the 4-dimensional one [4, p. 141].

By filling out the details I have omitted, one would have a complete spin foam model. However, several motivations make the quantum gravity community suspect that a spin foam model built on a fixed 2-complex is not the final theory of quantum gravity they were aiming for. In particular, one would like to consider arbitrary 2-complexes and define path integrals that sum over all possible 2-complexes compatible with a boundary geometry. Some heuristic reasonings support this idea, like spin foam expansions of LQG models, or the intuition that a continuum limit needs to be taken that takes into account finer and finer triangulations.[6] Be it as it may, GFT is precisely an approach that gives a prescription of how to take into account this sum over 2-complexes.

The key observation motivating the introduction of GFT is the realization that expression (1) is very similar to the one that gives the amplitude associated with a given diagram in a quantum field theory, which is usually given by a product of some amplitudes associated to each interaction vertex and some propagators which join the different vertices. In the case of the spin foam we find the same structure and it is tempting to interpret the spin foam as a Feynman diagram.[7] And if Feynman diagrams arise in perturbative expansions of QFTs, then spin foams should also arise from expansions of some theory. Group field theories are precisely the mathematical structures that are such that when treated perturbatively give rise to spin foam amplitudes.

In the amplitudes of a spin foam model the fundamental variables are the holonomies *h*, while in a standard Feynman diagram the basic variables are the positions and momenta of a set of particles, which are seen as excitations of a field. That the excitations of the field are characterized by positions is because the field itself is a field which is defined over position space.[8] Similarly, if we think of holonomies as excitations of some field, it needs to be a field that is defined on some group manifold, since holonomies are group elements. Thus, we arrive at the conclusion that if spin foams are Feynman diagrams of an underlying field theory, this field theory needs to be a group field theory.

For instance, let's consider the Boulatov model [9], which I will show is related to gravity in 3 spacetime dimensions. The field $\varphi$ in this model is defined on a group manifold consisting of three copies of SU(2). In other words, $\varphi$ is a function that takes three elements of SU(2) and gives a real number. The action for such a field is given by:

$$S_B[\varphi] = \frac{1}{2} \int \mathrm{d}^3 g \varphi^2(g_1, g_2, g_3) \\ - \frac{\lambda}{4!} \int \mathrm{d}^6 g \varphi(g_1, g_2, g_3) \varphi(g_1, g_4, g_5) \varphi(g_2, g_5, g_6) \varphi(g_3, g_6, g_4). \tag{2}$$

---

[6] See [4] for a discussion of these different motivations and intuitions. According to the perspective of these authors, considering all possible 2-complexes or a continuum limit allows one to achieve more precision, just as considering more diagrams of a Feynman expansion in a QFT allows for obtaining more accurate results. However, Dittrich has argued [6, 7] that the need to find a continuum limit is not just a matter of accuracy, as it could fundamentally change our picture of the theory and as it could be needed to preserve the diffeomorphism invariance of general relativity. I am thankful to an anonymous reviewer for highlighting these different perspectives on the continuum limit to me.

[7] This observation was first made in Ref. [8].

[8] Similarly, we can think of the field as defined in momentum space and the particles to be characterized by their momenta.

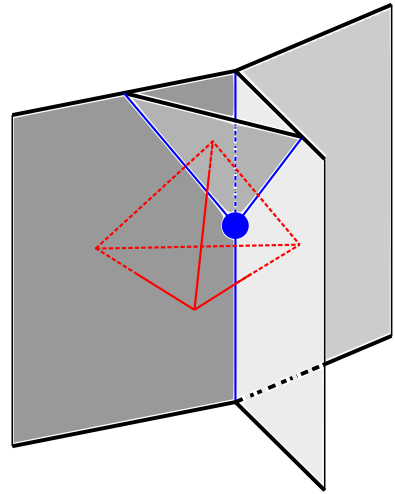

**Fig. 1** Section of a 2-complex in 3 dimensions which is dual to a tetrahedron (in red). We see that in the center of the tetrahedron there is one vertex (blue dot) and that there are four edges (in blue) going out from it, which are dual to the four triangles of the tetrahedron. The 2-complex has 6 faces (in different shades of grey), which are dual to the 6 segments of the tetrahedron. Notice that each edge of the 2-complex intersects just one triangle and that each segment of the tetrahedron intersects just one face of the 2-complex

**Table 1** Correspondence between elements of a 2-complex and its boundary and dual triangulations in 3 and 4 dimensions

| 2-complex | Triangulation (3D) | Triangulation (4D) |
| --- | --- | --- |
| Vertex | Tetrahedron | 4-simplex |
| Edge | Triangle | Tetrahedron |
| Face | Segment | Triangle |
| Node (boundary) | Triangle | Tetrahedron |
| Link (boundary) | Segment | Triangle |

We can define a 'partition function' as a path integral for such an action and expand perturbatively on the interaction parameter $\lambda$ to get:

$$Z = \int \mathcal{D}\varphi e^{-S_B[\varphi]} = \sum_\Gamma \lambda^{V(\Gamma)} \sum_{j_f} \prod_f d_{j_f} \prod_v \{6j\}_v = \sum_\Gamma \lambda^{V(\Gamma)} Z_{PR}(\Gamma). \tag{3}$$

The expansion in an interaction parameter in a QFT gives rise to a sum that can be represented by means of Feynman diagrams: each diagram represents a term in the sum and is multiplied by as many powers of the interaction parameter as interaction vertices in the diagram. In the expression for the Boulatov model the expansion can also be represented diagramatically as in Figs. 2 and 3. Each term in the sum is labeled by a graph $\Gamma$ which is formed by triple strands which meet in groups of four at vertices[9] and form a closed graph. Now, we can see each loop in this graph as defining a surface, and this, together with the structure[10] of the diagram defines a 2-complex, which is dual to a simplicial decomposition of a closed three-dimensional space. In particular, each strand is associated with an edge of the triangulation, each triple of strands is dual to a triangle, and each vertex is dual to a tetrahedron, as it is the place where 4 triangles meet. In Table 2 I summarize the relations between the parts of the diagrams, the 2-complex, and the dual triangulation.

Moreover, these diagrams carry irreducible representations $j$ of the group SU(2) and we find that the amplitude associated with each of these graphs is essentially a product of 6$j$ symbols, which is just a function of the six representations associated with the different strands meeting at the vertex. One can check that the amplitudes associated with these graphs correspond[11] to the amplitudes of the Ponzano-Regge model, a spin foam model in three dimensions. In this sense we find what we were expecting: the Feynman diagrams of a GFT correspond to spin foams, and the GFT expansion gives us a prescription of how to sum over different spin foams.

This duality can be explored further and for instance we can build quantities like the following[12]:

$$K[f] = \int \mathcal{D}\varphi f(\varphi) e^{-S_B[\varphi]}, \quad (4)$$

where $f$ is any functional of the field which is invariant under SU(2) gauge transformations of the form $\varphi(g_1, g_2, g_3) \to \varphi(hg_1, hg_2, hg_3)$ for any $h \in$ SU(2). The reason why this gauge restriction is imposed on the GFT is that it will be translated to the gauge restriction of the dual spin foam model. In particular, a basis in momentum space for the space of functions that satisfy this condition can be shown to be a set of spin network functions, i.e., a family of functions found in LQG and associated with spin networks. We can label these functions by a graph $\Gamma$ and an assignment

---

[9] The reason for this representation is that fields have three arguments, and hence the triple strand. The interaction term is a product of four fields, and hence four excitations meet at each vertex and combine following the same pattern of the interaction term $\varphi(g_1, g_2, g_3)\varphi(g_1, g_4, g_5)\varphi(g_2, g_5, g_6)\varphi(g_3, g_6, g_4)$ as can be seen in Fig. 2.

[10] By this I mean that the three surfaces associated with a triple strand share an edge, and each vertex of the diagram is a vertex of the 2-complex, as it is the place where 6 surfaces meet.

[11] There is a small issue here with the definition of the model, as the usual definition of the model carries some sign factors that don't appear in expression (3). I am taking here the expression from [10] which is the one discussed in the context of GFT. In Ref. [11] it is explained that the right expression for the Ponzano-Regge action is the one with the sign factors and points out that many times in the literature the wrong expression is given. For our discussion here this will not matter, as this example illustrates the point that in a perturbative expansion of a GFT there appear amplitudes of spin foam models.

[12] Here I am following the construction in Ref. [12, Sect. 9.3.3] but adapted to the case of the Boulatov model.

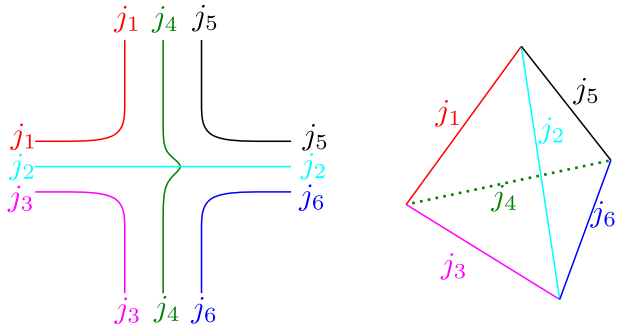

**Fig. 2** Interaction vertex of the Boulatov model in the *j* representation and the tetrahedron which can be interpreted to be represented by such an interaction vertex

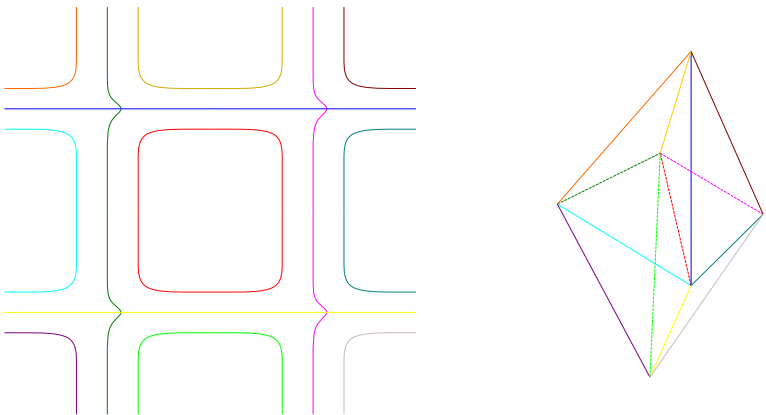

**Fig. 3** Portion of a Feynman diagram of the Boulatov model which corresponds to four tetrahedra glued together to form an octahedron. The closed red line in the diagram corresponds to the dual of the red edge in the octahedron, i.e., the line (not drawn) that goes around the red edge joining the four tetrahedra

of representations $j$ to edges of the network.[13] The field $\varphi$ takes three arguments, which implies that the spin network functions in this case are restricted to graphs with nodes of valence 3. For these functions we can perturbatively expand the quantity I have just defined to find:

$$K(s) = \int \mathcal{D}\varphi f_s(\varphi) e^{-S_B[\varphi]} = \sum_{\Gamma_s} \lambda^{V(\Gamma_s)} \sum_{j_f} \prod_f d_{j_f} \prod_v \{6j\}_v = \sum_{\Gamma_s} \lambda^{V(\Gamma_s)} K_{PR}(\Gamma_s). \tag{5}$$

---

[13] For models in higher dimensions spin network functions are also labeled by intertwiners at the nodes of the graph. Notice also that the graph here is purely combinatorial, as it arises from the contractions of polynomials of $\varphi$. In this sense, this spin network space is smaller than the original Hilbert space of LQG, which included networks that allowed for knotting and that included other diffeomorphism invariant information, the moduli.

**Table 2** Correspondence between the elements of the Feynman diagrams of the Boulatov model and the different parts of the 2-complex and dual triangulation. The relation between the diagram and the triangulation should be clear from the representation in Figs. 2 and 3

| Feynman diagram | 2-complex | Triangulation |
| --- | --- | --- |
| Vertex | Vertex | Tetrahedron |
| Triple strand | Edge | Triangle |
| Single strand (closed loop) | Face | Edge |

This expansion is the same that we found for the partition function but with one important difference. Now the sum is over graphs $\Gamma_s$ which correspond to 2-complexes that have the spin network $s$ as a boundary. $K_{PR}(\Gamma_s)$ corresponds to the amplitude that the Ponzano-Regge model[14] assigns to a 2-complex (defined by $\Gamma$) bounded by a spin network $s$. In this sense, we have obtained what we were looking for, i.e., a prescription of how to compute objects like $K(s)$ which isn't limited to just some fixed 2-complex, contrary to what happened in spin foam expressions of the form (1).

This example can be generalized to other spin foam models and in more spacetime dimensions by considering different GFTs, that is, by considering field theories defined on different group manifolds and with different actions.[15] For instance, generalizations of the Boulatov model like the Ooguri model[16] give rise to diagrams like the one in Fig. 4 and are associated with gravity in 4 dimensions. In this case the strands are quadruple, each single line represents a triangle, and each set of four lines, a tetrahedron. I express the correspondences between the diagrams of this model, the 2-complex and dual triangulation in Table 3.

As we have just seen, GFTs can be seen as a way of defining a prescription for computing transition amplitudes $K(s)$ in a systematic way that goes beyond considering a fixed 2-complex. However, GFT is considered by its practitioners as a theory on its own and not just as a useful computational tool for LQG or some discrete approach. In the rest of the paper I will analyze the way GFT has been interpreted as an independent theory which also describes an emergent spacetime.

I will also argue that there are two different ways in which GFT has been interpreted as a theory on its own. The first one is the covariant one which takes the theory to be defined by means of the expressions of the partition function and expectation value I have just introduced, while the second one uses the canonical formalism of quantum mechanics. These two ways are presented in the literature as equivalent[17], but I will argue against this claim. I will further argue that neither of them is satisfactory. It is unclear in which sense the covariant formulation constitutes a quantum theory, and it seems to amount just to a computational method for computing spin foam amplitudes. In this sense, it doesn't offer an improvement over them, and it may be subject to the same conceptual shortcomings. On the other hand, the canonical approach seems to be ill-motivated, and it suffers from a problem of time that in my opinion makes the approach unsatisfactory.

---

[14] Again, up to some sign corrections, as explained in footnote 11.
[15] This relation between GFTs and spin foam models was established in Ref. [8].
[16] See for instance [13, Sect. 2.1] for an introduction to this model and the diagrams it produces. The original formulation of the model can be found in Ref. [14].
[17] The distinction between the two is not commonly made explicit. I refer the reader to [15, Sect. 2.2] for an example of how the canonical version of GFT is introduced as a reformulation of the covariant theory.

## 2 Covariant GFT

The covariant view of GFT is the one arising from the structures I have introduced in the previous section, namely, a definition of a path integral over field configurations on a group manifold. In this section I argue that it is unclear that this construction of a path integral constitutes a quantum theory in any sense which is stronger than the claim that it defines a partition function $Z$ and a series of 'expectation values' $K(s)$. In this sense, I argue that it is doubtful that this view of GFT offers something over and above a prescription about how to sum over different spin foams. Moreover, one can worry about what justifies the form of the GFTs that are claimed to have something to do with quantum gravity in our world and that GFT doesn't solve the deep conceptual issues that the other approaches to quantum gravity face.

Covariant GFT is to be understood as a quantum field theory. However, let me start by analyzing the classical counterpart to GFT to highlight how different the formalism of GFT is from the formalism of a standard quantum field theory. A classical GFT would be defined by specifying a group manifold and an action that fields on this group manifold would have to minimize. In this sense, an action like the Boulatov action or the Ooguri action I have introduced in the previous section would define classical theories, in the sense of determining a preferred set of field configurations that would obey some classical 'laws' that would be obtained by varying the corresponding action. However, the analogies with classical field theories as we understand them end here. In the first place, the manifolds on which these field theories are defined are not spacetime manifolds and hence this classical theory is defined on some abstract space and we are missing a connection with our experience.

The other great difference is that the way one could conceive of a dynamics in the GFT case would be very different from the dynamics of a classical theory. The actions typically used in GFT are non-local from the point of view of the group manifold, as is explicit in the interaction term in Boulatov action (2). In a standard field theory the equations of motion of the theory are local, i.e., the value of a field at one point is determined by its value and possibly the value of other fields in a neighborhood of the point. Actions like (2), on the other hand, lead to non-local equations of motion, and the value of the field at a point of the group manifold depends not only on the points in its neighborhood, but also on points arbitrarily far.[18] This makes it the case that we cannot approach the classical GFT using many of the standard tools of classical mechanics which are later used for formulating a quantum theory, such as the Hamiltonian formalism.

---

[18] Notice also that, as far as it is possible to use spatiotemporal notions for this example, this non-locality is not restricted to be the spatial non-locality that there arguably is in quantum mechanics. Instead, this non-locality is global, as points everywhere in the manifold influence any given point. If the manifold were a spacetime we would say that the future and the past non-locality affect the present.

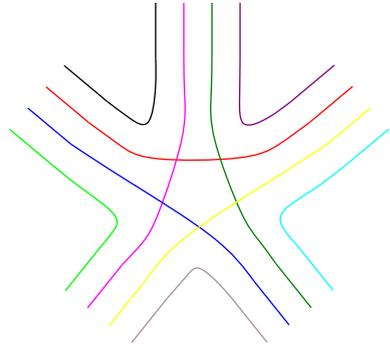

**Fig. 4** Interaction vertex of the Ooguri model. This vertex is dual to a 4-simplex in four dimensions

**Table 3** Correspondence between the elements of the Feynman diagrams of the Ooguri model and the different parts of the 2-complex and dual triangulation. Notice that the relation between the elements of the diagram and the 2-complex has not changed, while the elements of the triangulation are one dimension higher

| Feynman diagram | 2-complex | Triangulation |
| --- | --- | --- |
| Vertex | Vertex | 4-simplex |
| Quadruple strand | Edge | Tetrahedron |
| Single strand (closed loop) | Face | Triangle |

This analysis of the classical counterpart of GFTs raises the worry of how to define a quantum GFT. As just argued, it doesn't make sense to perform a canonical analysis of a classical GFT or to interpret the group manifold as a spacetime manifold, and hence the standard routes for canonical quantization are blocked. What is postulated by the GFT community is instead a 'covariant' route, that is to define the quantum version of the theory by means of some path integrals.[19] In particular one usually defines two kinds of objects, namely, partition functions and expectation values:

$$Z = \int \mathcal{D}\varphi e^{-S[\varphi]} \quad (6)$$

$$\langle f \rangle = \int \mathcal{D}\varphi f(\varphi) e^{-S[\varphi]}. \quad (7)$$

As I commented in the previous section, these expectation values can be interpreted as the amplitudes that one would obtain by summing over all possible spin foams with a boundary determined by some functional $f$ of the field. Now, are these definitions enough to claim that we have defined a quantum theory in a meaningful way? In the following I will argue for a negative answer.

---

[19] This definition of GFT was the one given in the earlier works on the approach. See [3, 16].

The first thing to notice is that, in coherence with my analysis above, the definition is not in terms of the standard structures of quantum theories, such as Hilbert spaces, operators, and wavefunctions. In this sense, if by quantum theory we mean something related to the evolution of a wavefunction, as seems to be the case in the most extended realist interpretations of the formalism, it seems that we do not have a quantum theory. However, one may try to claim that the definitions (6) and (7) implicitly define a quantum theory that relies on the standard structures. One can see the canonical version of GFT that I will analyze in the following section as a proposal of the canonical structures that would be so implicitly defined, but I will argue that this version of GFT is just postulated, and the way it is related with covariant GFT is doubtful. So far, in the literature this equivalence is, as far as I know, just postulated and no proof or hint for the equivalence is provided.

One could try to argue for the acceptability of expressions (6) and (7) as giving a definition of a quantum theory by relying on some claims and results in the quantum field literature that show that the content of a quantum field theory is encoded in certain expectation values. However, one needs to be cautious here, as there are two senses in which claims in this direction are made, and only one of them could be applicable to the case of GFT.

The first of the senses in which expectation values contain the information of a quantum field theory is by means of certain theorems that are able to relate certain vacuum expectation values with the transition amplitudes for arbitrary processes. In this sense, knowing the expectation values is equivalent to knowing the dynamics of the theory and one has everything needed for recovering the predictions of the theory and rebuilding the standard structures. This being true for QFTs, it is not applicable to the case of GFT. This is because these theorems rely on the canonical structures of quantum field theory, which are not available for the case of GFT.[20]

There is a second sense in which a quantum field theory can be seen to be contained in its expectation values. There is an approach to quantum field theory, axiomatic quantum field theory, that tries to define quantum field theory in terms of algebraic structures such as an algebra of observables and a state that would assign expectation values to each observable in the algebra.[21] Typically, these observables are associated with field observables at spacetime regions and it is aimed to recover the results of Lagrangian QFT in a mathematically more rigorous way. In a weak sense, covariant GFT can be seen as defining a quantum theory if we take this definition, as it defines an algebra of 'observables' (the functionals $f$) and a way of determining expectation values of them (using expression (7)). However, it is clear that there is a difference between both cases, as there is an intuitive meaning of the observables in the case of the QFT, while it is not so clear in the case of a GFT. In the case of QFT it is more or less straightforward to connect the formalism with the standard pictures of quantum mechanics

---

[20] These theorems are known as the LSZ theorems, and the point that they rely on the canonical formalism has been also made in Ref. [17].

[21] I refer the reader to [18, 19] for an introduction and defense of axiomatic field theory.

and quantum field theory while in the case of GFT we don't seem to have this connection.

For all these reasons, I believe that GFTs as defined by expressions (6) and (7), i.e., what I am referring to as covariant GFTs, only constitute a quantum theory in the weak sense of defining a set of observables and a way of computing expectation values for them, and not in a stronger sense that is needed for the standard ways of thinking about the quantum formalism. Even if we accept this, there are two remaining points that can be problematic from the point of view of the foundations of this approach.

If covariant GFT is reduced to a way of computing some quantities whose meaning is given by some other approach to quantum gravity such as LQG, then it seems that the role of GFT is not to be something like an independent theory that incorporates insights from different approaches, but just something similar to a computational tool with no physical interpretation. In this sense, one shouldn't try to understand the GFT as a theory or its action as something like defining a law of nature. Work in GFT can be useful for having well-developed techniques for computing the relevant objects, but it seems that in order to have fundamental progress one would need to make advances with some theory of quantum gravity for which we could have an interpretation, and not with a formalism for which no clear interpretation is given. In this sense, if the covariant approach to GFT aims to constitute a candidate theory for quantum gravity, it cannot limit itself to define some formalisms which agree with some other approaches in some aspects and it should also provide an interpretation.

Finally, even if one accepted this, there is another potential problem with the approach, namely, that it inherits the interpretational problems that spin foam models have. That is, if the goal of GFT is to compute amplitudes like $K(s)$, for it to be fully successful it needs to provide a story about how these objects relate to the world. There are some reasons to think that spinfoams models are problematic, as for instance they have a problem of time in that they do not define a time-evolving wavefunction and it is unclear how to think about these models using the standard ways of thinking about quantum mechanics. If you believe there is a problem with the interpretation of spin foam models, then GFTs do not seem to help much.

To conclude this section, let me insist that covariant GFTs are defined in terms of some expressions which make use of path integrals but in a way that seems to be unrelated to the quantum theories we are familiar with. In this sense, they seem to be reduced to useful computational techniques for obtaining quantities like $K(s)$ and they do not offer an improvement in the foundations of quantum gravity beyond giving a prescription about how to sum different diagrams.

## 3 Canonical GFT

As I have argued in the previous section, it is doubtful that the path integrals introduced in Sect. 1 define a quantum theory that relies on the canonical structures of quantum theories such as Hilbert spaces, operators, unitarily evolving wavefunctions, and so on. Moreover, the classical GFTs aren't defined on spacetime manifolds and they are non-local in a strong sense, which forbids writing them in the Hamiltonian formalism and applying canonical quantization methods. However, in the GFT community a canonical quantum theory motivated by the ideas of GFT has been proposed. As I have argued above, this sort of approach is just postulated and not derived from the covariant formulation and there are reasons to doubt that the two approaches are in some sense equivalent. In any case, in this section I will introduce this approach, which I call canonical GFT, as it relies on the canonical structures of quantum mechanics, and I will comment on its shortcomings, namely that its motivation and justification can be challenged and that it presents certain conceptual troubles related to the problem of time of canonical approaches to quantum gravity.

The way canonical GFT is postulated resembles other approaches to quantum gravity, in which there is an auxiliary Hilbert space known as the kinematical Hilbert space which is used for defining dynamically relevant states as states which satisfy some dynamical condition. The space formed by these states is known as the physical Hilbert space and states in this space are supposed to represent the physical content of the theory. In the case of GFT one also defines an auxiliary Hilbert space, which I analyze in Sect. 3.1, and then imposes a dynamics on it, which I analyze in Sect. 3.2. Let me emphasize that, while in canonical approaches to quantum gravity this particular structure is a consequence of the application of the canonical quantization procedure to gauge systems like general relativity, in the case of GFT this structure is many times just postulated. In this sense, the canonical formalism of many GFT models is unrelated to the standard canonical quantization procedure, as we will see below.[22]

### 3.1 The Kinematical Hilbert Space

The Hilbert space postulated[23] for GFT is the Fock space for the field defined on the group manifold of the GFT we are considering. This is motivated by considering the field as what plays a fundamental role in the theory, but there is no further justification for this, given that we are not following any quantization procedure but just directly postulating a model. The Fock space for the GFT model is however an interesting one, as it has some resemblances with the Hilbert space of LQG.

Recall that Fock spaces are the spaces used in QFT for quantizing a field and allow for a particle interpretation, since one can decompose a Fock space as a sum over subspaces that correspond to a different number of field excitations or particles:

$$\mathcal{F} = \bigoplus_n S(\mathcal{H}^{\otimes n}). \tag{8}$$

Here $\mathcal{H}$ represents the Hilbert space of one such particle or excitation. In the case of a QFT like a scalar field theory this corresponds to the space $L^2[\mathbb{R}^3, dx]$, that is, the Hilbert space used for describing the quantum mechanics of one particle in three-dimensional space. The space $\mathcal{H}^{\otimes n}$ is the product of $n$ copies of this space, and hence represents a state which describes $n$ particles living in space.

---

[22] I will also comment on a class of models in which a canonical quantization procedure is followed by adding a new variable to the model which is effectively treated as a time variable.
[23] See for instance the way canonical GFT is presented in Ref. [20].

To completely define a Fock space one also needs to specify the symmetry conditions that will be satisfied by the states of the theory, which is imposed by means of the symmetry operator $S$, which restricts the symmetry properties of the subspaces of the space. In the case of GFT it is generally chosen[24] to have symmetric states, which allow for the bosonic statistics needed for defining condensate states which are used in cosmological models inspired by GFT.[25]

Let me remark on an important difference between the Fock quantization of a QFT and the quantization carried out in GFT. In the case of the QFT, what one quantizes is the state of the field at a given time, that is, the Fock space represents the possible particle content of the world at a time slice, and the time evolution of a state in this space represents how particles move, interact and possibly create and annihilate.[26] Meanwhile, the Fock quantization of GFT is not associated with something like a time slice but it is defined for the whole group manifold. This already shows that the way dynamics is defined will be different, as the QFT picture of a state evolving representing the particles at different time slices won't be available. Furthermore, it raises a worry about the strength of the analogy of GFT with QFT, as, as far as I know, in the context of QFT no Fock space is built in such a global way.

The most remarkable feature of the Fock space defined in GFT is that it can be given a particle-like interpretation, but the space in which particles live is not ordinary space, but the group manifold of the theory. In this sense, the role that position played in characterizing a QFT state is now played by the group variables of the theory. In the case of GFTs aiming to recover a 4-dimensional spacetime one natural option[27] is to choose the field to be defined on four copies of SU(2). This means that the single-particle Hilbert space is chosen to be $L^2[SU(2)^4]$, which has as a basis the states $|g_1 g_2 g_3 g_4\rangle$, characterized by the four group elements $g_1$, $g_2$, $g_3$, and $g_4$. For a state representing $n$ particles, the Hilbert space is $L^2[SU(2)^{4n}]$. This Hilbert space is related to the Hilbert space of LQG as I will discuss now.

A way of defining the kinematical Hilbert space structure of LQG is as a direct sum of graph Hilbert spaces $\mathcal{H}_\Gamma$:

$$\mathcal{H} = \bigoplus_\Gamma \mathcal{H}_\Gamma \tag{9}$$

---

[24] In the case of a QFT defined on spacetime the spin-statistics theorem makes it the case that depending on the spin of the particle one symmetry condition or the other needs to be applied, leading to the two different kinds of particles: bosons and fermions. As GFTs are not defined on a spacetime there is more freedom at the time of defining the theory in the most convenient way.

[25] These cosmological models are outside the scope of this paper and I refer the interested reader to [1, 21–24] for some of these models and reviews.

[26] There are many subtleties involved at the time of defining and interpreting a QFT, including but not excluding issues regarding whether states should be interpreted as associated to an instant of a foliation of spacetime or they should be treated in a more covariant way. For the discussion in this section we can leave these subtleties aside and take just the general simplified picture I have just discussed.

[27] It allows to make a connection with approaches like LQG and some spin foam models.

Each graph Hilbert space is characterized by a graph, and states in these spaces are roughly interpreted as representing a discrete space made of chunks in which each node in the graph represents a chunk and each link an adjacency relation. For a graph with $L$ links, the Hilbert space is $L^2[SU(2)^L]$. This Hilbert space is very similar to the $n$-particle Hilbert space of GFT $L^2[SU(2)^{4n}]$, and suggests a relation between both approaches. Indeed, it has been suggested to take the GFT quanta as representing 4-valent nodes of a spin network (that is as nodes with 4 incoming or outcoming links). A general GFT state would represent an open graph in which nodes are disconnected from each other, while there would be a subset of the GFT states that would correspond to 'glued' nodes that would form a closed graph and this would be just equivalent to LQG states according to the proponents of GFT. The precise details of how these 'glued' states can be defined and identified can be found in Ref. [25].

However, we have strong reasons for believing that the relationship between the Hilbert spaces of GFT and LQG is merely heuristic and that the GFT gluing is problematic. First, the gluing procedure defined in the GFT context has been argued[28] to be ill-defined in the sense that a single 'glued' state does not correspond to a unique graph, but to several. Therefore, we have one state in GFT which corresponds to several different networks with different topological and geometrical properties and which are represented by different states in LQG. Second, notice that nodes in GFT are restricted to be four-valent, as the one particle space was defined on four copies of SU(2), while in LQG we have nodes with arbitrary valence, i.e., connected with an arbitrary number of nodes. Besides, networks in GFT are purely combinatorial, while in LQG the networks one finds by applying the canonical quantization procedure have additional structure, namely knotting classes and moduli. In the LQG literature there are different views about how to interpret this structure and a debate about whether it is necessary for the theory.[29] If one believed these structures to be fundamental, then GFT would fall short, as it is unable to represent them. For all these reasons, the connection between GFT and LQG seems to be just heuristic.

GFT states can be also connected with triangulations of space similar to the ones I have introduced in section 1. If one imposes some gauge condition[30] on the one-particle Hilbert space above, one obtains states of the form $|j_1, j_2, j_3, j_4, v\rangle$ which can be given a geometric interpretation. Any such state would represent a tetrahedron of volume $v$ and the area of each of its faces would be a function of the corresponding $j_i$. This seems to suggest an identification of GFT states (when glued, and assuming for the sake of argument that this could be done) with the boundaries of approaches to quantum gravity based on triangulations or spin foams.

However, one also needs to be cautious about making such an identification. In the first place, not every assignment of quantum numbers gives rise to a geometry consistent with a triangulation of a 3-dimensional space. That is, some values of the *j*'s and *v*'s are compatible only with more general geometries known as twisted geometries[31] and that aren't formed by flat tetrahedra. In the second place, not every way of gluing tetrahedra together is compatible with a triangulation of space. In this sense, not every state in the GFT Hilbert space is compatible with the triangulation picture.

---

[28] See the argument in section 6 of [26].

[29] In Refs. [27–29] one can find arguments for and against considering these structures as worth keeping in LQG.

[30] This condition is that the field is required to obey some symmetry condition like being invariant under transformations of the form $\varphi(g_1, g_2, g_3, g_4) \to \varphi(hg_1, hg_2, hg_3, hg_4)$ for any $h \in SU(2)$.

Finally, let me comment on how GFT has been mentioned in debates about the emergence of spacetime. Indeed one could argue that the way GFT states relate with a 3-dimensional space[32] is one of emergence, as we arguably have a case in which a space is in some sense approximated or constituted by very different structures, as presumably the group-field-theoretic entities are. However, one also needs to be careful with this sort of claim, as was noted for instance in Ref. [31]. The reason for this is that depending on how we interpret the basic theory, the GFT in this case, the claim of emergence may be trivial or much harder to prove than expected. As Dowker and Butterfield note, if one takes the discrete parts of a discrete approach to quantum gravity to be literally chunks of space, then the claim of emergence is easy and very weak. That is, if we interpret the GFT particles to be tetrahedra then we are already assuming a lot of geometric properties and it is no surprise that we are able to recover a space. On the other hand, if one takes the particles to correspond to something much more abstract, then it remains a challenge to close the gap and provide a story about how the structures of GFT relate to space. In this sense, from a functionalist perspective as defended in Ref. [32] one needs to relate the abstract GFT state with the functions we usually attribute to space.

To sum up, let me insist that the kinematical Hilbert space of GFT is postulated to be a Fock space and that particles in this space are interpreted as something like chunks of space, as there are some similarities with LQG and some discrete approaches. However, I have noted that there are some differences between them, and in the following I will argue that they suffer from similar conceptual issues at the time of formulating the dynamics.

### 3.2 GFT Dynamics

The way the dynamics is defined in this canonical version of GFT is not by specifying a Hamiltonian that would define a time evolution for the discrete spaces that one can arguably associate with some GFT states. Instead, the dynamics is defined by postulating that physically meaningful states are the ones that satisfy certain conditions, just as in the case of theories like LQG physically meaningful states are the ones that satisfy certain constraints. I will now introduce how these conditions are implemented in GFT and the challenges that this way of defining the dynamics faces.

To make a connection with covariant GFT, the condition that 'physical' states have to satisfy is defined is related to one of the actions I introduced in Sect. 1. In particular, the constraint equation which is imposed is the following:

---

[31] See [30] for an introduction to the notion of twisted geometry.
[32] Here I am referring to the models considered in this section. More generic models could be related with spaces of any dimensionality.

$$\frac{\delta S[\hat{\varphi}]}{\delta \hat{\varphi}(g)}|\psi\rangle = 0, \tag{10}$$

where we have quantized the classical action by replacing the fields with field operators.[33] In Ref. [15] it is argued that this choice of dynamical equation is a natural one: by varying the action with respect to the field one obtains the classical equation of motion, by translating this equation to one expressed using quantum operators[34] we obtain a natural candidate for the constraint. Another alternative is to select the states such that they satisfy a tower of equations known as the Schwinger-Dyson equations:

$$\langle\psi|\hat{O}\frac{\delta S[\hat{\varphi}]}{\delta \hat{\varphi}(g)}|\psi\rangle = \langle\psi|\frac{\delta \hat{O}}{\delta \hat{\varphi}(g)}S[\hat{\varphi}]|\psi|\rangle. \tag{11}$$

These equations have to be satisfied for every operator $\hat{O}$ in the GFT Hilbert space. In particular, notice that a state satisfying (10) automatically satisfies the Schwinger-Dyson equation for $\hat{O} = \mathbb{I}$. In this sense both choices are approximately equivalent. The option of selecting states that satisfy Schwinger-Dyson equations is of extended use in the GFT literature.[35]

In standard QFTs, Schwinger-Dyson equations like the ones I have presented arise naturally as a consequence of functional calculus applied to path integrals. In the case of GFT a similar motivation has been given,[36] by using path integrals like the ones defined by (6) and (7). However, as I have argued in the previous section, these expressions are not clearly associated with a Hilbert space structure, and hence it is not possible to identify expressions of the form $\langle\psi|\hat{O}|\psi\rangle$ (like the ones in 11) with expressions of the form $\langle\hat{O}\rangle$ as defined by the path integral (7). In this sense, one cannot use the machinery that one uses in the case of QFT to justify these equations, and hence the connection of the dynamics as defined in the canonical version to GFT with the covariant version remains heuristic. The imposition of Eq. (11) seems a natural choice but it is important to note that it is not derived in any principled way.

This analysis raises the worry that the canonical version of GFT isn't fully justified, as the relationship with the covariant version of GFT and the motivations for adopting it seem to be merely heuristic. A different route for the justification of this form of the dynamics was provided in Ref. [34]. There, it was shown that one can derive an equation like (10) by starting with the Hamiltonian constraint of LQG and translating it into a constraint in the Hilbert space of GFT, which can be done given the relation between the Hilbert spaces of both approaches that I have highlighted above. In this sense, one can try to justify the adoption of canonical GFT by appealing to LQG. However, one may still worry that we have very weak reasons to adopt canonical GFT, as everything we have is a heuristic connection with covariant GFT (which in Sect. 2 I also argued that was problematic) and a relationship with LQG. If GFT is reduced to some alternative version of LQG or a slightly changed version of it, then one may worry that canonical GFT doesn't offer an improvement over LQG.

---

[33] As is typical in quantum mechanics there may be some ordering ambiguity in this step.
[34] As emphasized at several points of this thesis this process is not trivial, as there may be ordering ambiguities.
[35] See for instance [20, 33].
[36] See for instance [34, Sect. 5]

Let me make a technical remark. In canonical quantization programs, the imposition of the constraints leads to a definition of a new Hilbert space, the physical Hilbert space, which is formed by the states that satisfy the constraints. Typically, one also needs to define a new inner product in this Hilbert space, given that the inner product of the original, unconstrained Hilbert space is singular between states that satisfy the constraints. Despite this being a well-known feature of canonical quantization, it is largely overlooked in the GFT literature,[37] where it is usually the case that the constraints are imposed but there is no definition of a new inner product. This issue could lead to unwanted divergences in the formalism.

To conclude this section I will argue that, when adopting the way the dynamics is defined in LQG and canonical approaches to quantum gravity, canonical GFT also adopts one of the conceptual problems of these approaches, namely a problem of time. This problem arises from the fact that one doesn't define quantum states which evolve with respect to some time parameter[38] but just claims that states in the physical Hilbert space contain all dynamical information. The challenge is therefore to recover a picture with a time evolution from states in this physical Hilbert space. For approaches like quantum geometrodynamics, LQG, and also canonical GFT, states in this space are superpositions of 3-geometries, possibly discrete and with some caveats to consider, and there isn't any explicit time variable in these Hilbert spaces. There have been several proposals for how to interpret these timeless superpositions of geometries as something from which our intuitive picture of a time evolution could be recovered, but they all face important conceptual and technical challenges that make them not convincing from my point of view. I refer the reader to [37–41] for some introductions and criticisms of the proposed resolutions of the problem of time. In the case of canonical approaches to GFT, and especially in the cosmological models inspired by it, the way the problem of time is dealt with is by means of a relational strategy. In the following I, argue that this approach is problematic.

Relational strategies for solving the problem of time of canonical approaches to quantum gravity essentially consist of taking one of the degrees of freedom of the model and effectively treating it as a time variable. For instance, if the canonical quantization procedure leads to a wavefunction $\psi(x, y, z)$ which we would naively interpret as a superposition of three physical degrees of freedom represented by $x$, $y$, $z$, the relational strategy is to interpret the state as describing how the quantum state of two of these variables evolves with respect to the third. That is, to interpret the wavefunction as a state $\psi(x, y)$ evolving in $z$ for instance.

In my opinion, when $x$, $y$, $z$ are physical degrees of freedom such as the positions of particles or the values of some field it is conceptually wrong to treat one of them as a time variable.[39] The reason for this is that one is conflating the concept of time with the concept of clock. It is true that we use physical systems to keep track of how other systems evolve, but this does not mean that we should treat physical variables describing clocks as we treat time variables or that the temporal structures in our theories (and in the world) could be eliminated.[40] This conceptual confusion leads to several problems and issues at the time of applying this strategy for solving the problem of time. First, we can see that there is some asymmetry in the way $x$, $y$, $z$ are interpreted in the quantum theory: while $x$ and $y$ are quantum and subject to quantum phenomena $z$ is taken to be a time variable and it won't experience superpositions, entanglement, or other phenomena. Second, there is a multiple choice problem as one can build different theories by picking up a different 'clock'. Moreover, these theories could be radically different. For instance, in the cosmological context it has been argued [45, 46] that changing the variable one chooses as a clock changes whether the theory predicts a singularity or not and whether one has a universe that expands indefinitely or not. Third, for some models there may be no variable that could work as a clock. Fourth, the relation with the classical theory is dubious and it is unclear how to recover it in some approximate manner.

---

[37] A couple of exceptions are the articles [35, 36], where an inner product for the physical Hilbert space is indeed defined to be different from the original one.

[38] Here I am using the Schrödinger picture for describing the problem, but the problem arises also if we use the Heisenberg picture or any alternative representation of the standard dynamics of a quantum theory.

I refer the reader to [41] where I develop these points in more detail and to [37–40] and references therein for other critical views. Studying the GFT case allows highlighting some of the general worries one may have with relational resolutions, besides some specific issues that appear in the GFT context.

For being able to apply a relational strategy in GFT one customarily adds at least one variable to the formalism so that it can be used as a clock. In this sense, the GFT is now defined to be a field $\varphi(g, \phi^0, ..., \phi^N)$ depending on the group variables $g$ and some other ones $\phi^0, ..., \phi^N$. The action is also modified to include this dependence on the new variables. The new form of the action is argued [47] to reproduce the transition amplitudes of spin foam models that include some scalar fields, and hence the variables $\phi^i$ are claimed to represent scalar fields.

Now the relational strategies for solving the problem of time take one of the fields, say $\phi^0$ as a 'clock'. I will mention a couple of ways in which this idea has been implemented in the GFT literature. Notice that different authors differ about whether adding more fields is necessary. On the one hand, those with a relationalist intuition [47] argue that three more fields are necessary for being able to distinguish spacetime points and analyze non-symmetric spaces. On the other hand, we have argued that the GFT states are similar to LQG states in that one can argue that they describe geometries of space, and these can be highly inhomogeneous. In this sense, the motivation from a relationalist perspective is weakened and in many works in GFT one introduces just one scalar field. For the rest of this section I will consider just the addition of one field and focus on the issue of time.

---

[39] The confusion arises because this kind of system is many times compared with parametrized models. In the case of parametrized models one defines wavefunctions on an extended configuration space in which one of the variables is a time coordinate (or several variables are spacetime coordinates). For instance, instead of having $x$, $y$, $z$ one has $x$, $y$, $t$. In this case it is perfectly fine to deparametrize and interpret this wavefunction as a state for $x$, $y$ evolving in $t$. But when the model one is considering is not defined on such an expanded configuration space, one cannot apply the same reasoning. I expand on this distinction on [41].

[40] I refer the reader to [42–44] for critical discussions of radical versions of relationalism motivated by quantum gravity and which aim to eliminate time from our theories.

A common way to implement the relational idea in GFT is to enlarge the Hilbert space of canonical approaches to GFT by building a bigger Fock space, one in which each particle carries information about both the group and field variables. The straightforward interpretation of such a Hilbert space would be just the same as before but with the addition that now the discrete geometries carry more information. The relational resolution of the problem of time seeks to identify a clock variable and defines time evolution as evolution with respect to this variable. In this case, we would like this clock variable to be the field $\phi^0$ but we find the problem that $\phi^0$ is not directly an observable in the GFT Hilbert space. Instead, each particle has associated its own value of $\phi^0$, so it is like every particle is carrying its own clock. In this sense, if we insist on interpreting $\phi^0$ as measuring time, generic states describe particles that are at different 'times' and cannot generally be decomposed in terms of a nice temporal evolution. That this Hilbert space is problematic for defining relational dynamics is recognized and discussed in more detail in Ref. [48].

In cosmological applications of GFT[41], the states considered are restricted to condensate states, which present certain symmetries that allow to avoid the above problems and a relational dynamics can be defined. For the rest of states it seems that the case of GFT is not different from the one of other theories with a problem of time: once we consider realistic situations with some variety of variables involved, it becomes hard to define a relational dynamics in a way that allows to formally overcome the problem of time. And, from a conceptual perspective, the idea of having to identify a clock or a time variable in a context in which there are several which are as good or as bad as any other seems to be begging the question. For instance, consider that a three-particle state $\psi(g_a, \phi_a^0, g_b, \phi_b^0, g_c, \phi_c^0)$ was a solution of the modified version of (10). Certainly, choosing either of the three 'clocks' $\phi_i^0$ as the time variable that defines the dynamics seems ad hoc and unjustified, as the three of them would seem equally good or equally bad. This sort of problem raises important worries about the feasibility of the relational resolution, and the worry extends also to the symmetric cases that are used in cosmological models, even though the symmetry hides these conceptual issues.

More recently, part of the GFT community has taken a different relational way of dealing with the problem of time [35, 36, 50]. In Ref. [50] it is shown how one can treat the scalar field variable $\phi^0$[42] as a time variable even before applying any quantization procedure. Then, the author performs a classical canonical analysis taking $\phi^0$ to be the time variable and quantizes the resulting theory. This theory is just a regular quantum theory describing how a quantum state evolves with respect to the time variable $\phi^0$, and it avoids some of the technical issues that arise in the other approach discussed above. Alternatively, in Refs. [35, 36] an equivalent way of arriving at the same theory is developed based on arguably more covariant techniques which make use of the constrained formalism.

One can take this canonical approach to exemplify the way the concepts of time and clock are mixed in relational strategies for solving the problem of time. Now it is not only that $\phi^0$ is treated in the quantum theory as a time variable, but also that it is treated as a time variable in the classical setting. In this sense, one has recovered all the structures of classical mechanics, i.e., there is a configuration space, a phase space, a Lagrangian, and a Hamiltonian. The only difference is that the time variable instead of being called $t$ it is called $\phi^0$ and that there were some arguments that connected it with a scalar field. For this reason, one can suspect that what one is effectively doing in this context is to add a temporal structure to a theory that had none in order to make sense of it. As I commented above, the way we treat physical variables such as scalar fields in both classical and quantum theories is radically different from the way we treat temporal structures, and in the models discussed in Refs. [35, 36, 50] it seems clear to me that $\phi^0$ is treated as a temporal structure rather than as a physical field.

---

[41] See the recent reviews in Refs. [1, 49].

[42] $\chi$ in their notation.

This, together with the general worries discussed above make the relational strategies implemented in GFT conceptually problematic in my opinion. To sum up, the canonical versions of GFT suffer from a problem of time which makes it vulnerable to the same kind of objections that can make us think that approaches like geometrodynamics or LQG are untenable or that we haven't found the right way to think about them yet. If we add to this the worry that the canonical versions of GFT seem to be just heuristically connected with the covariant one and with the motivations that led to introducing GFT in the first place, we are naturally led towards a skepticism about this approach, at least while the worries raised here concerning the justification and interpretation of the approach have been addressed.

## 4 Conclusions

In this paper I have offered an introduction to GFT where I have raised several concerns that one may have regarding this approach to quantum gravity. GFT presents interesting connections with other approaches such as spin foam models or LQG but I have argued that there are several issues to be solved in order to clarify the status of GFT and what it could say about quantum gravity.

In the first place, I have noted that there are two ways of presenting GFT which are only heuristically connected. I have argued that the covariant definition of GFT, i.e., the definition in terms of path integrals, seems to be just limited to a way of computing certain 'expectation values' with no connection with a quantum theory as we usually think about. These expectation values correspond to some spin foams transition amplitudes, and hence covariant GFT can be seen as a way of computing these objects. However, this would leave covariant GFT as a very limited approach and vulnerable to the same objections that one can raise against this class of models.

Then, I have introduced the canonical version of GFT. This version is of extended use in the cosmological models inspired by GFT and hence it is particularly relevant. As I have argued, the disconnection of the canonical and covariant versions of GFT makes it the case that we lack a motivation for the canonical version of GFT beyond some heuristic connection and an inexact correspondence with LQG. Furthermore, I have argued that this approach suffers from a problem of time and that the relational ways proposed for dealing with it have an ad hoc feel, and that they suffer from the well-known technical and conceptual shortcomings that all relationalist approaches share.

**Acknowledgements** I want to thank the Proteus group, Carl Hoefer, and Daniele Oriti for their comments and discussions. I also want to thank an anonymous reviewer for their useful comments. This research is part of the Proteus project that has received funding from the European Research Council (ERC) under the Horizon 2020 research and innovation programme (Grant agreement No. 758145) and of the project CHRONOS (PID2019-108762GB-I00) of the Spanish Ministry of Science and Innovation. Open Access funding provided thanks to the CRUE-CSIC agreement with Springer.

**Funding** Open Access funding provided thanks to the CRUE-CSIC agreement with Springer Nature. This research is part of the Proteus project that has received funding from the European Research Council (ERC) under the Horizon 2020 research and innovation programme (Grant agreement No. 758145) and of the project CHRONOS (PID2019-108762GB-I00) of the Spanish Ministry of Science and Innovation.

**Data Availability** NA.

**Code Availability** NA.

**Declarations**

**Conflict of interest** The authors have no Conflict of interest to declare that are relevant to the content of this article.

**Ethical Approval** NA.

**Consent to Participate** NA.

**Consent for Publication** NA.